\documentclass[12pt]{article}

\usepackage{setspace} 
\usepackage[margin=1truein]{geometry}

\usepackage{amssymb,amsfonts,amsmath}

\newcommand{\mxan}{\textit{M. xanthus }}

\title{Phase transitions during fruiting body formation in \textit{\textbf{Myxococcus xanthus}}}

\author
{Shashi Thutupalli,$^{1,2,3}$ Mingzhai Sun,$^{2}$ Filiz Bunyak$^{4}$, \\Kannappan Palaniappan$^{4}$, Joshua. W. Shaevitz$^{1,2,5}$\\
\\
\small{$^{1}$Joseph Henry Laboratories of Physics, Princeton University, Princeton, NJ 08544, USA}\\
\small{$^{2}$Lewis-Sigler Institute for Integrative Genomics, Princeton University, NJ 08544, USA}\\
\small{$^{3}$Department of Mechanical and Aerospace Engineering, Princeton University, Princeton, NJ 08544, USA}\\
\small{$^{4}$Department of Computer Science, University of Missouri, Columbia, Missouri, 65211, USA}\\
\small{$^{5}$email: shaevitz@princeton.edu}
}

\date{\today}

\usepackage{graphicx}

\begin{document}
\maketitle

\begin{abstract}
The formation of a collectively moving group benefits individuals within a population in a variety of ways such as ultra-sensitivity to perturbation, collective modes of feeding, and protection from environmental stress. While some collective groups use a single organizing principle, others can dynamically shift the behavior of the group by modifying the interaction rules at the individual level. The surface-dwelling bacterium \emph{Myxococcus xanthus} forms dynamic collective groups both to feed on prey and to aggregate during times of starvation. The latter behavior, termed fruiting-body formation, involves a complex, coordinated series of density changes that ultimately lead to three-dimensional aggregates comprising hundreds of thousands of cells and spores. This multi-step developmental process most likely involves several different single-celled behaviors as the population condenses from a loose, two-dimensional sheet to a three-dimensional mound. Here, we use high-resolution microscopy and computer vision software to spatiotemporally track the motion of thousands of individuals during the initial stages of fruiting body formation. We find that a combination of cell-contact-mediated alignment and internal timing mechanisms drive a phase transition from exploratory flocking, in which cell groups move rapidly and coherently over long distances, to a reversal-mediated localization into streams, which act as slow-spreading, quasi-one-dimensional nematic fluids. These observations lead us to an active liquid crystal description of the myxobacterial development cycle.
\end{abstract}

\section{Introduction}

The collective motion of individuals that exhibit complicated group dynamics is a hallmark of living systems from single-celled bacteria to large mammals. Collective groups can gain advantages including ultra-sensitivity to perturbations, increased temporal response, and increased protection from the environment~\cite{Cavagna2010,Zitterbart2011,Ancel1997}. It is often considered that individuals follow simple interaction rules that give rise to surprising group phenomena as an emergent property~\cite{Moussaid2011, Herman1979}. However, in many cases, how a group decides to change its behavior, or alternatively how groups can perform multiple functions, remains unclear~\cite{Parrish1999}. Do individuals have to perform increasingly more complicated tasks, or can they merely transition between a preset number of simple interaction rules to modify group behavior~\cite{Zhang2012}?

A striking manifestation of multicellular collective behavior is the formation of  dynamic cells groups by the soil-dwelling bacterium \emph{Myxococcus xanthus} \cite{Zusman2007, Peruani2012, Starruss2012}. In a plentiful environment, the coherent motion of cells allows them to hunt prey  through the cooperative production of antibiotics and digestive enzymes \cite{Rosenberg1977}. In contrast, if a swarm cannot find sufficient nutrients, its cells begin a complex, multi-step process that leads to the formation of giant aggregates called fruiting bodies within which many of the cells sporulate \cite{Zusman2007}. This process takes several ($\sim$ 12-72) hours and involves multiple distinct stages of group behavior \cite{Shimkets1990}.

Fruiting body formation is at its heart a process of cell density coarsening and aggregation. However, unlike many classical aggregation phenomena where a reduction in mobility gives rise to static aggregates as the motion of the individuals becomes essentially frozen, \mxan cells and cell groups remain  dynamic throughout the process, often moving over long distances as  fruiting bodies are born, grow and coalesce. These cells are able to amass cell groups that retain both cellular and group mobility even as the density coarsens over time. For this reason, models that invoke a density-dependent reduction in cell speed fail to capture subtleties of  \mxan group dynamics. For example, Sliusarenko et al. observed a  reduction in cell speed  during the early stage of fruiting body formation when cells approached nucleation regions of high local cell density~\cite{Sliusarenko2007}. These observations supported a model in which cells form aggregates passively due to jamming inside  dense cell clusters. However, this model yields a frustrated aggregate that cannot perform the dynamic group motions often seen later in fruiting body development. A more complicated interaction rule must be at play.

A common feature of biological collective motion across different length scales is its similarity to the flow of fluids. In particular, the orientational ordering within these collectives is reminiscent of liquid crystalline fluids~\cite{Marchetti2013, Adhyapak2013}. Active nematic fluids~\cite{Marchetti2013,Ramaswamy2010}, such as a bird flock or a collection of \mxan cells~\cite{Peruani2012}, exhibit giant number fluctuations (GNFs) in two or more dimensions~\cite{Simha2002}. These density changes, where the standard deviation of the number of individuals in a group of size $n$ grows faster than $\sqrt{n}$, promote the disintegration of cell aggregates. While it is clear that fruiting body development must remain active throughout the entire process, dynamic aggregation can be inherently unstable making the formation of localized aggregates difficult; how \mxan overcomes this issue is unknown.

Perhaps the most striking feature of \emph{M. xanthus} motility is periodic directional reversals that drive unique collective modes such as wave-like dynamic ripples \cite{Shimkets1982, Mauriello2010}. Myxo cells, which glide in the direction of their long axis, possess a dynamic cell polarity that routinely switches direction by 180 degrees \cite{Zusman2007}. These reversal events are accompanied by the exchange of a number of polarity and motility proteins between the leading or lagging poles which then switch roles. Reversals are  critical for complex group behavior, a link first observed by Blackhart \cite{Blackhart1985}. Without a properly functioning reversal mechanism, \emph{M. xanthus} cells fail to order themselves within a swarm, and non-reversing cells are unable to produce an expanding swarm \cite{Wu2009}. 

A key limitation of previous studies of \mxan collective behavior is that they were based on observing either large cell groups at low optical magnification incapable of resolving single cell behavior, or only a small number of individuals within a large group where cell-cell dynamic interactions could not be studied. In most of these studies, only a small number of cells are analyzed, either manually or semi-automatically, due to a lack of robust algorithms for tracking thousands of cells simultaneously. Recent automated approaches have been limited in scaling up to produce dense long term cell tracks~\cite{AlberChen2014}.The small size of these datasets precludes an in-depth statistical study of group behavior.

\begin{figure}[htbp]
   \centering
        \includegraphics[width=\textwidth]{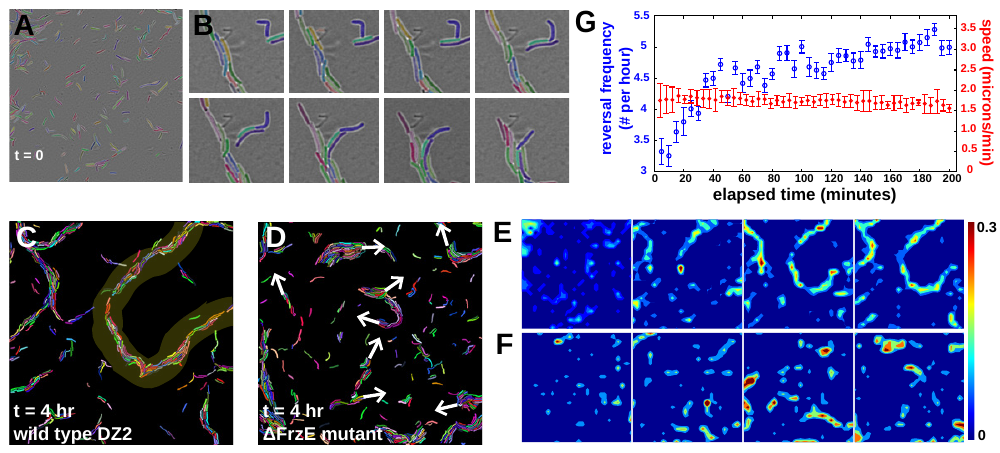}
    \caption{Changes in local cell density and reversal frequency of wild type \textit{Myxocococcus xanthus} cells affect their aggregation patterns. (\textbf{A}) A typical dispersion of a monolayer of \textit{M. xanthus} cells on a feature-less and nutrient-less surface of agarose gel at the start of an experiment. Individual cells are identified and marked to aid visualization. (\textbf{B}) Steric interactions between individual moving bacterial cells align their directions and cause linear aggregates to form over time. The sequence of images is shown every 40 seconds. (\textbf{C}) Over a duration of 4 hours after starvation, the local bacterial density in the aggregates increases to about $\approx$~ 1 cell$/3 \mu$m$^2$ for the wild type cells. The growth of the aggregates of wild type cells is confined to preferred locations in space, which eventually form stream-like elongated aggregates. (\textbf{D}) Non-reversing, $\Delta$FrzE cells form coherently moving flocks, groups of cells that all move in the same direction (arrows). (\textbf{E, F}) Cell density graphs (cells/$\mu$m$^2$) shown at hourly intervals for the wild type (E) and non-reversing mutant (F). When the ability of the bacteria to reverse is suppressed, the aggregates of the non-reversing mutant are highly mobile and grow and shrink dynamically without forming streams. Scale: All individual panels are 110 $\mu$m X 110 $\mu$m. (\textbf{G}) The reversal frequency (blue) and speed (red) of the wild type cells over time.}
    \label{trackingfig}
\end{figure}

In order to develop a statistical model of \mxan group dynamics, we developed a high-throughput computational image analysis platform to measure the position and motion of each cell in a population of thousands of cells over several hours to bridge the dynamics of single cell motion with the emergence of mesoscale group order. Here we focus on understanding the very initial stages of fruiting body formation, when sporadically distributed cells find each other to form clusters that merge and grow into larger cell groups. We show that reversal frequency, instead of cell speed, is the key factor that regulates the group behaviors of cells which switch from an initial flock-like searching to static aggregation. Nematic alignment and cellular reversal produce one-dimensional, stream-like aggregates that are not subject to GNFs.

\section{Results and Discussion}


\subsection{Cell tracking in densely packed groups}

We imaged the motion of cells directly after starvation on an agar pad every 10 seconds for four hours. To track the motion of individual cells at high local spatial densities, we developed the custom-written BCTracker bio-image informatics software (Materials and Methods) that automatically segments dense cells and tracks these cells over time. This algorithm is able to track all $\sim 1,000$ cells per movie, including those in large, densely packed groups. Overall, we detected, segmented and tracked nearly 4 million individual bacteria organized into more than 44 thousand filtered tracks across three four-hour long movies for both the wild type {\em M. xanthus} DZ2 strain and the reversal-deficient mutant $\Delta$FrzE~\cite{Li2005} respectively. This resulted in a total of 2,159,196 and 1,763,993 tracked cells respectively.

\subsection{Changes in reversal frequency and local cell density cause cells to aggregate into streams}

In our experiments, wild type bacteria are randomly oriented and distributed at the onset of starvation as seen in Fig.~\ref{trackingfig}\textbf{A}. These isolated cells find each other via a series of contacts and collisions as they glide over the substrate (Fig.~\ref{trackingfig}\textbf{B}).  Steric interactions, combined with collisions, lead to the local alignment of the cells whereby neighboring cells in a group are all aligned in the same direction. These local nematic aggregates of cells move together over the course of a few hours. As the bacteria start to aggregate through these collisions, the local density of  cells increases about 10 fold from $\sim$~0.1 cell~$/$~3$\mu$m$^2$ to $\sim$~1 cell~$/$~3$\mu$m$^2$ within the aggregates. For the wild type bacteria, this increase in density is localized spatially with very dense regions surrounded by large voids (Fig.~\ref{trackingfig}\textbf{C, E}, Movie S1). The combination of an elongated shape, nematic ordering, and a high density ultimately leads  to the formation of stationary, stream-like aggregations of cells (Fig.~\ref{trackingfig}\textbf{E}). 

In order to evaluate the effect of directional reversals on this aggregation behavior, we used a FrzE mutant strain which lacks the ability of  cells to reverse. In these non-reversing cells, the local density of the cells increases at a similar rate to the wild type cells (Fig.~\ref{trackingfig}\textbf{F}). However, the aggregates that result are no longer elongated, stream-like, or stationary. Instead, the $\Delta$FrzE mutants form flocks of collectively moving cells that appear as motile high density blobs in Fig.~\ref{trackingfig}\textbf{D, F} and Movie S2. Comparing this to the wild type cells,  cells in the streams do not move cohesively in the same direction due to their periodic reversals. As a result, the streams are largely fixed in space and increase in width over time as more bacteria align and join the stream.

Over the course of the experiment, we observed a marked increase in the reversal frequency of the cells whereas the cell speed remained constant during all four hours of observation (Fig.~\ref{trackingfig}\textbf{G}).  During the first hour after starvation, the reversal frequency of the cells rapidly increased from 3 reversals per hour to 5 reversals per hour (Fig.~\ref{trackingfig}\textbf{G}). During this period however, the speed of the bacteria remains almost constant at $\sim$~1.5~$\mu \rm m/min$. Taken together, these observations suggest that in the first hour after starvation, cells move persistently in a certain direction for a greater distance than at later times, allowing them to efficiently explore space and search for neighbors.

\begin{figure}[htbp]
   \centering
        \includegraphics[width=\textwidth]{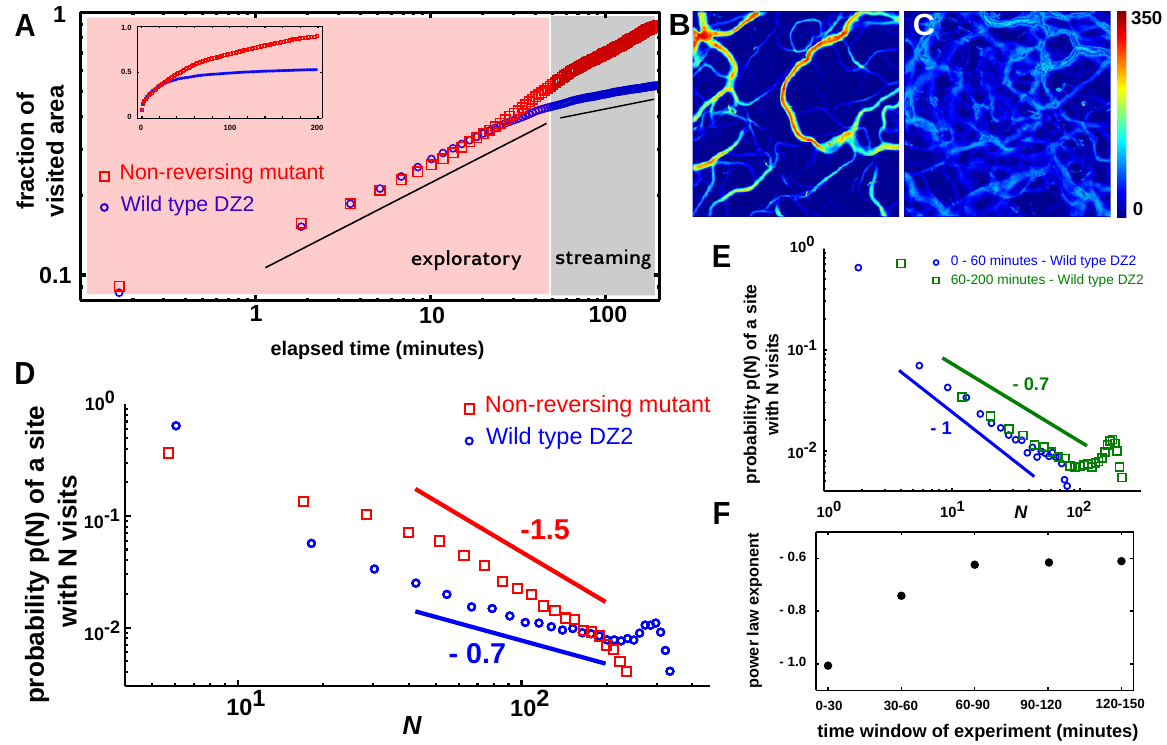}
    \caption{Spatial exploration by the bacteria is arrested due to the aggregation into streams and the aggregation is marked by a transition from an exploratory flocking phase to the quasi-stationary streaming phase. (\textbf{A}) The dynamics of the spatial exploration by the cells. Black lines are a guide to the eye to mark the change in the slope during the 'exploratory' and 'streaming' phases. The inset shows the same data on a linear scale to show the saturation of the visited area fraction for the wild type cells. A map of the visit frequencies for (\textbf{B}) the wild type DZ2 cells during the first 4 hours following starvation and (\textbf{C}) non-reversing $\Delta$FrzE cells. (\textbf{D}) The probability distribution of the number of visits \textbf{$N$} of a given site in space by a cell follows a power law decay for the wild-type and $\Delta$FrzE mutants. The appearance of the peak in the number of visits is seen for the wild-type cells. (\textbf{E}) The site visit probability for the wild type cells in the first 60 minutes and last 140 minutes of a 4 hour long experiment. (\textbf{F}) The power law exponent of the decay of the probability distribution (calculated from the dynamics of the cell motions in 30 minute windows of the experiments).}
    \label{explorfig}
\end{figure}

\subsection{Spatiotemporal dynamics of aggregation and stream formation}

The temporal dynamics of the exploration of space by the cells is quantified by (i) the rate at which the bacteria visit the various regions in space and (ii) the frequency with which they repeat such visits. Fig.~\ref{explorfig}\textbf{A} shows the fraction of the total available area that is visited by the bacteria. While the wild type cells do not explore more than half of the available area, the non-reversing mutants explore nearly the entire field of view. We can characterize the temporal dynamics of this spatial exploration into two phases: an 'exploratory phase' and a `streaming phase'. In the  `exploratory phase', exhibited by wild type cells during the first hour after starvation and non-reversing cells throughput the experiment, cells rapidly explore space, allowing them to search for nutrients and to initiate aggregation by finding neighboring bacteria. An hour after starvation,  wild type cells undergo a phase transition from the flocking phase to the streaming phase. This slows down the exploration of space dramatically as the cells remain localized within the streams. The non-reversing mutants, however, continue to flock  in groups over the entire four-hour experiment.

To quantify how the motion of individual cells was distributed across space, we constructed a surface-visit map for all the bacteria in a field-of-view, i.e. we probed the number of times a location in the field of view had been visited by a bacterium (see Materials and Methods). We find that a considerable amount of the space remains unexplored for the wild type  bacteria (Fig.~\ref{explorfig}\textbf{B}) and the highest number of visits are localized to the region in space that eventually forms the stream. For the non-reversing $\Delta$FrzE mutant cells, however, the visits are distributed more evenly across space (Fig.~\ref{explorfig}\textbf{C}). To combine the information from multiple experiments, we calculated the probability distribution, $p(N)$, of the number of pixels with  $N$ visits from the surface-visit maps (Fig.~\ref{explorfig}\textbf{D}). Over a range of values of $N$, this distribution exhibits a power law decay. While the decay exponent is about $-1.5$ for the $\Delta$FrzE mutant, it is less steep, an exponent of $-0.7$, for the wild type cells.

This power-law decay reveals features of the motility mechanism. Any feature of the system that causes bacteria to have memory of a site they, or another cell, previously visited will result in a bias in the way space is explored when compared to an unbiased random walk. This bias causes some locations to be more preferable than others and may be a result of the interaction with slime trails deposited on the substrate~\cite{Burchard1982}. In addition, the steeper slope of the decay for $\Delta$FrzE  cells indicates that they explore space more uniformly than the wild type as is evident also from the visit map in Fig.~\ref{explorfig}\textbf{B}. 

Strikingly, while $p(N)$ decreases monotonically with $N$ for the non-reversing bacteria, the distribution for the wild type cells is marked by a peak in the probability at higher values of $N$. This peak indicates a significant probability that a few sites are preferentially visited many times. This occurs because of the formation of localized streams in which the reversing bacteria traverse over the same sites over and over.

The  distribution $p(N)$ changes during the first hour of aggregation, consistent with the occurrence of a phase transition. When calculated using data from the first hour, the power-law exponent is $-1$ and does not display the peak at large $N$. In contrast, the distribution from subsequent times has a slope of $-0.7$ and contains a prominent peak at $N\sim200$ (Fig.~\ref{explorfig}\textbf{E}). Furthermore, the power law exponent of the decay gradually increases over time from the flock-like value to the stream-like value (Fig.~\ref{explorfig}\textbf{F}). As time proceeds, the number of preferred locations for localization of the bacteria increases (i.e. along the stream), each drawing a large number of visits due to the reversal of the cells. 


\begin{figure}[hbtp]
   \centering
        \includegraphics[width=\textwidth]{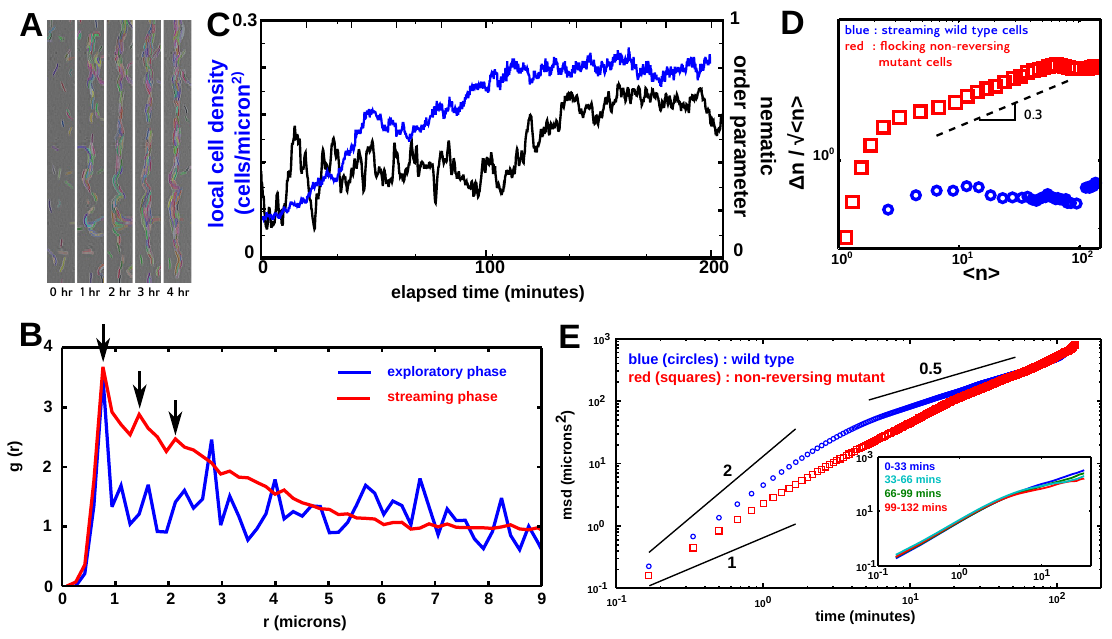}
    \caption{The stream-like aggregates are a quasi one dimensional active nematic. (\textbf{A}) A series of images shows the nematic like ordering of the cells along the long axis of the stream as it forms over time. (\textbf{B}) The radial distribution function $g(r)$ for the bacteria in the stream region during the exploratory and the streaming phases. Data is analysed in the exploratory and streaming phases for bacteria from 5 different streams, each comprising a few hundred bacterial cells. (\textbf{C}) As the local cell density increases, the cells align along their body axis, leading to an increase in the local nematic order. The nematic order is an average over 5 streams, each containing a few hundred cells, from 3 different experiments. (\textbf{D}) Number fluctuations for the flocking non-reversing mutants (red) and the wild type streams (blue). Number fluctuations are normal for the streams while they are anomalous (giant) for the flocks. (\textbf{E}) The mean squared displacement (MSD) as shown for the DZ2 (blue), which makes a transition from a ballistic motion at the short time scales to an anomalous sub-diffusive behavior at longer times. In contrast, the $\Delta$FrzE mutant (red) cells remains superdiffusive throughout. The MSD is an ensemble average of a few hundred cells for each of the cell types. Inset: The mean squared displacement is calculated for every 30 minute time slot from the start of the experiment.}
    \label{streamfig}
\end{figure}

\subsection{Streams act as one-dimensional active nematic highways}

Bacteria in wild-type streams are ordered much like rod-like molecules in a nematic liquid where the molecules orient themselves in a direction along their long axis. In this ordered state, the bacteria maintain liquid-like mobility through active motion and yet remain confined. This allows them not only to navigate eventually into a fruiting body but also prevents the streams from breaking apart. Fig.~\ref{streamfig}\textbf{A} shows the formation of a stream in a region where the bacteria are initially oriented in random directions. As these bacteria move and collide with one another, the steric interaction during the collision provides a simple physical mechanism to orient them along their body axes such that they lie parallel to each other. Once aligned in a particular direction, the bacteria then move along that direction leading to a new set of collision and realignment events with other bacteria. 

The overall effect of such dynamics is the formation of a stream with all the bacteria oriented along the stream axis. During this process, the bacteria make a phase transition from an isotropic gas-like phase to a nematic liquid like phase in the streams. This can be seen from the radial distribution function $g(r)$ in Fig.~\ref{streamfig}\textbf{B}. The appearance of distinct  peaks in  $g(r)$ for the streaming phase  shows  the liquid-like ordering of the bacteria within the stream. This liquid-like ordering also has an additional orientational ordering of the bacteria due to their elongated shape. This ordering can be quantified using an order parameter $Q=\sqrt{<{\rm cos~2}\theta>^2 + <{\rm sin~2}\theta>^2}$, where $\theta$ is the angle between the body axis of the bacteria and a direction of reference oriented along the direction of the stream. Limiting cases correspond to $Q=0$ for a perfectly disordered state and $Q=1$ when all the bacteria are perfectly aligned with their neighbors. Initially, the random orientation of the bacteria results in a low value of the order parameter $Q \sim 0$ and as the stream builds up to reach a steady density, the ordering dynamics progressively orient the bacteria, leading to a nematic state within the stream (Fig.~\ref{streamfig}\textbf{C}). 

Active nematics in two dimensions or more are marked by the presence of  giant number fluctuations (GNFs)  where the standard deviation $\Delta n$ of a mean number $n$ of active apolar particles grows faster than $\sqrt{n}$~\cite{Simha2002}. In general, this should lead to disruption of any aggregated groups and would be counter-productive for fruiting-body formation. These anomalous fluctuations can be compared to more common fluctuations where the standard deviation $\Delta n$ grows as $\sqrt{n}$ in accordance with the central limit theorem. In this case, there are no abnormally large fluctuations in the density of conventional system undergoing common fluctuations whereas density is not a well-defined quantity in an active nematic system.

GNFs, however, are suppressed in one dimension so that $\Delta n \sim {n}^{0.5}$. This is what we observe for  bacteria in  streams~\cite[Fig. \ref{streamfig}\textbf{D} blue]{Simha2002}. In contrast, the dynamic flocking behavior of the non-reversing cells leads to anomalous giant fluctuations  such that $\Delta n \sim {n}^{0.5+0.3}$ as reported previously for bacteria ~\cite[Fig. \ref{streamfig}\textbf{D} red]{Peruani2012, Zhang2010}. Therefore, even while the wild type bacteria remain active and motile in the stream, this motility does not lead to large fluctuations in the bacterial density. This is presumably important for cells to maintain contact with each other and ensure continuity of the aggregates.

Motion within a stream is quasi one-dimensional, like cars along a highway. However, even though the bodies are aligned, their velocities remain uncorrelated due to reversals, i.e.  the nematic alignment is apolar without a single direction of motion. This allows the stream to remain fixed in space while maintaining mobility and avoiding GNFs. This high level of mobility  within the streams is revealed by examining the mean-squared displacement (MSD) as a function of time for the bacteria (Fig.~\ref{streamfig}\textbf{E}). At short times $t$, the MSD increases $\propto t^2$ due to the quasi one-dimensional ballistic motion of the cells. Surprisingly, at longer times, the MSD increases $\propto t^{0.5}$ indicating a sub-diffusive, constrained motion. This is due to the confinement of the cell motion in the quasi-one dimensional streams and the periodic reversals of the cell movement direction along the streams leading to head-to-head collision events~\cite{Richards1977}. In contrast, the MSD for the non-reversing mutants remains super-diffusive (i.e. MSD $\propto t^\alpha$, $\alpha > 1$) for all times. Indeed, this is expected given that these cells form flocks which traverse space in a persistent random walk. The anomalously slow (sub-diffusive) growth of the MSD for the wild-type cells indicates a very slow dispersal of the cells at long times which effectively prevents the loss of localization while still keeping the cells motile. 

\begin{figure}[htbp]
   \centering
        \includegraphics[width=\textwidth]{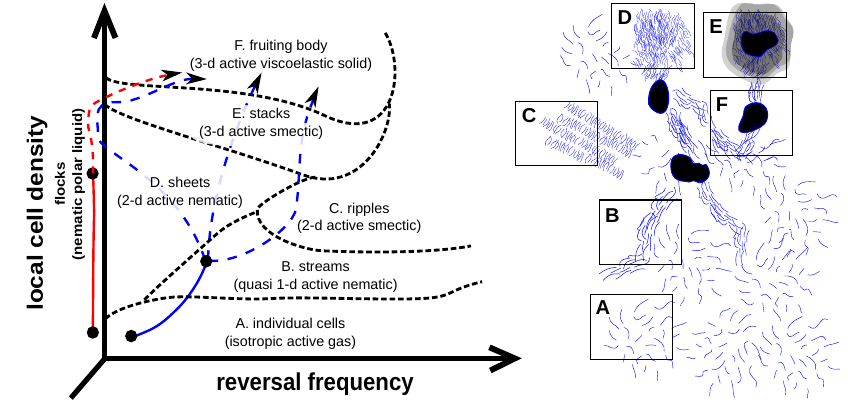}
    \caption{An active ordered fluids framework for the developmental cycle of \textit{Myxocococcus xanthus}. In a high-dimensional phase space, we consider cell density and reversal frequency of the Myxobacterial cells. The developmental program of the bacteria is marked by various ordered states such as (\textbf{A}) isolated cells: an isotropic active gas phase, (\textbf{B}) streams: a quasi one-dimensional active nematic fluid (\textbf{C}) ripples: in which the cells are ordered in rows like in a smectic liquid crystalline phase, (\textbf{D}) sheets: in which the cells are ordered in a 2 dimensional apolar active nematic fluid, (\textbf{E}) stacks: in which multiple sheets are ordered in smectic-like layers on top of each other and (\textbf{F}) fruiting bodies: an active viscoelastic solid phase. The blue and red lines represent the putative paths in phase space followed by wild type cells and non-reversing mutants respectively. The data for the solid lines is presented in this paper.}
    \label{phasediag}
\end{figure}

\subsection{An active fluids framework for {\em \normalsize Myxococcus xanthus} development}

The similarity of the collective motion of \mxan cells to the dynamics of fluids, in particular to liquid crystalline fluids, naturally suggests an underlying self-organization principle. The various structural aggregates of the \mxan from the isolated cells to the streams and fruiting bodies can be likened to the phase ordering in liquid crystalline fluids, with the switching between the different phases marked by well defined phase transition points. The key difference between biological systems and everyday fluids is that living matter is active; the individuals within the group are self-propelled. In statistical mechanics based theories and simulations, self-propulsion together with simple physical interactions between individuals, such as collisions and steric hindrance, has been shown to lead to collective motion phases and patterns bearing a striking similarity to natural phenomena~\cite{Marchetti2013, Ramaswamy2010}. The developmental cycle of the \textit{M xanthus} can be treated as a collection of various active fluid phase behaviors (Fig.~\ref{phasediag}) embedded in a high dimensional phase space involving both physical and biochemical effects. Our focus in this work is a subset of this phase space involving the changes in local cell density and the reversal frequency of the \mxan cells, in which the randomly organised cells at the very onset of starvation lie close to the lower left corner and can be treated as an isotropic active gas. The fruiting bodies that result from the aggregation of the bacteria are soft mounds which are a viscoelastic solid-like phase. The route from the isotropic gas phase to this viscoelastic solid is marked by phases of different levels of ordering and relevant to the developmental cycle of the \textit{M. xanthus}. As we have shown, the aggregation of the bacteria leads to a nematic liquid which manifests as quasi one-dimensional streams and polar flocks. Some of these active liquid phases and transitions between them have been shown to occur in purely physical systems such as systems of self-propelled rods~\cite{Peruani2012}. However, biological activity, in addition to self-propulsion, involves chemical communication between individuals and further downstream biochemical regulation of motility and interactions between individuals. The transition between these initial stages of organization of the \mxan is presumably coordinated by the motility of the cells, their physical interactions, as well as biochemical signaling between them~\cite{Pathak2012,Kim1990} and internal regulation within individuals. We have here shown, for example, that the regulation of the reversal frequencies of the cells governs one such transition. Future work defining the phases (both in 2 and 3 dimensions) and phase transitions explored by \mxan cells during predation and development, as well as the chemical and biological mechanisms that govern the control of individuals should lead to a much more in-depth understanding of how collective groups can exhibit multiple behaviors.

\section{Materials and Methods}

\subsection{Experiments and Tracking}

Myxococcus xanthus strains were grown in CYE medium (1 Casitone, 0.5 yeast extract, 10 mM 3-(N-morpholino) propanesulfonic acid (MOPS), pH 7.6, 4 mM MgSO4) overnight at 32oC to OD550~0.6. For \emph{M. xanthus} development, cells were washed in TPM (10 mM Tris-HCl, pH 7.6, 1 mM KH2PO4, 8 mM MgSO4) three times to remove residue nutrients from CYE medium. 2 l of cell solution were spotted on a 1 ultra-pure agarose pad prepared on a glass cover slide (1 ultra-pure agarose dissolved in TPM medium). A glass coverslip was then covered on top of the cells. 

The cells were imaged on a modified Nikon TE2000 inverted microscope with a 100x oil immersion objective (NA 1.49) using partially crossed-polarizer illumination. An image of the sample was projected onto a EMCCD camera yielding and effective pixel size of 85 nm and a total field of view of 43.5x43.5 $\mu$m. To increase the number of cells in out field of view, we employed a tiling strategy where a 3x3 grid of images was recorded every 10 seconds that were later post-prossesed into one large 110x110 $\mu$m image. 

To remove drift during  prolonged time-sequence imaging, we developed an image-based active feedback system. A $z$-stack of images of the central tile was taken and the sum of the Laplacian of the images was calculated as a focus index. This parameter is a strong function of the $z$-position and has a maximum at the highest contrast. This same auto-focus procedure was also done for every single tile at the beginning of the experiment to determine the relative best-focus position for each tile, using the central tile as the reference. Focus correction for all tiles of the image was then performed using the offsets recorded at the beginning of the experiment. 

\begin{figure}[htbp]
\centering
  \includegraphics[width=\textwidth]{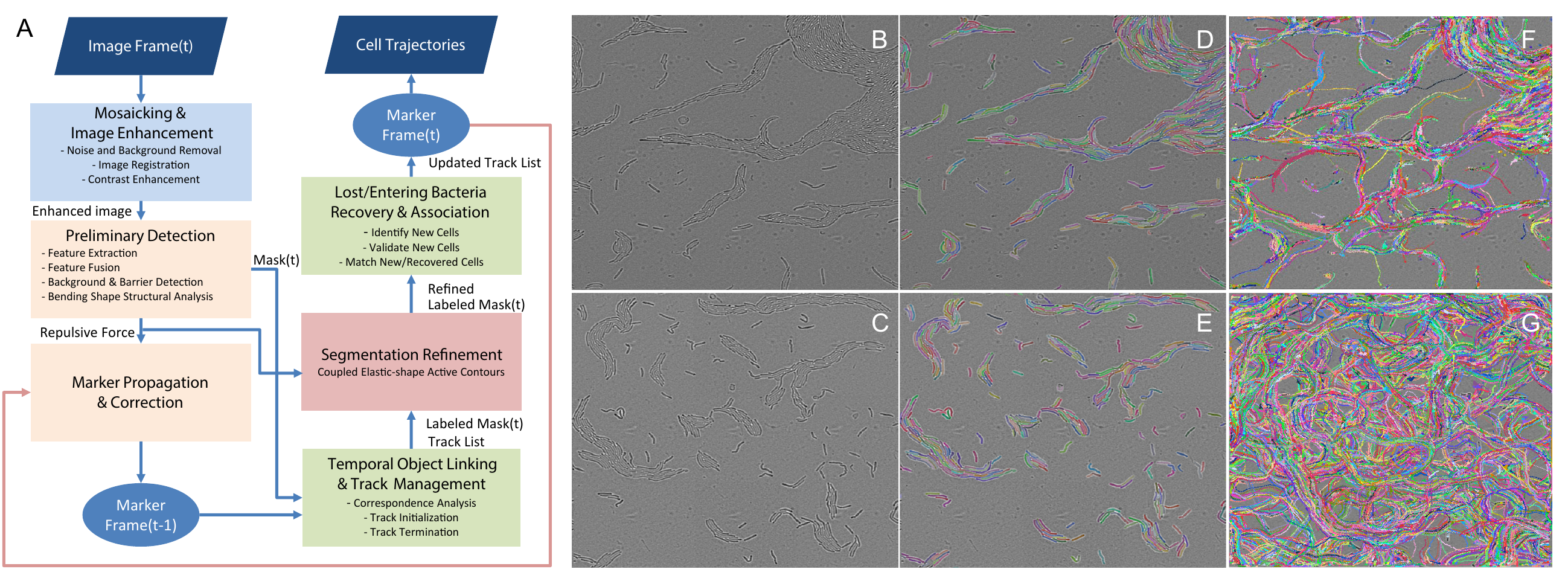}
  \caption{\textbf{A.} Flow chart of the BCTracker program. Images of bacteria are first preprocessed to reduce noise and enhance contrast. Candidate objects are detected using a preliminary
detection step, followed by image marker propagation and correction. Objects are linked
and segmentation is refined using a level-set-based active contour framework. \textbf{B-C.} Raw
microscope images of gliding M. xanthus cells. \textbf{D-E.} Segmentation results from the
images in B, C where different colors indicate individual bacteria. \textbf{F-G.} Overlay of all
trajectories from the entire movies from which still images \textbf{B, C} were taken. \textbf{B, D, F}:
wild type DZ2 strain. \textbf{C, E, G} $\Delta$~FrzE strain.
 }\label{BCfig}
\end{figure}

We have developed an automatic video analysis system BCTracker (Fig. \ref{BCfig}) for segmentation and long-term tracking of thousands of densely clustered individual bacteria expanding upon our previous work in image-based cell motility analysis~\cite{Bunyak:ISBI-2006, Palani:Raobook-2009, Ersoy:ISBI-2012-rbc}. The BCTracker high-throughput video analysis pipeline involves three major stages. First, the mosaicking and image enhancement module uses image-to-image registration for mosaic construction combined with several image restoration steps to compensate for illumination variation, increase contrast and filter out noise. Mosaicking is used to construct a larger field-of-view to support the accurate segmentation and tracking of larger bacteria cell groups as they undergo flocking or streaming behaviors where the groups can move between microscope fields, without sacrificing spatial resolution. Image enhancement improves subsequent detection, segmentation and tracking processes. In the second stage, the feature extraction and bacteria detection module is used to extract differential geometry and morphological image features tuned for the flexible rod shaped cells. Analysis of the extracted feature vectors results in a multivalued mask that identifies the foreground bacteria, background and halo regions around the cells. The multivalued mask contains positive and negative contextual regional information and is more versatile than a typical binary cell mask. An active contour energy function uses this multivalued mask and the fused feature set to evolve the contour adaptively to better locate, refine and segment the deformable thin rod shape of the cells more accurately. A structural analysis step uses spatiotemporal shape-based constraints like the bacteria skeleton or medial axis, and the neighborhood relationships to model the temporal interactions between spatially adjacent cells. The results of structural analysis combined with marker propagation in time and multi-frame evidence-based correction is essential for identifying and correctly segmenting touching cells. This is critical to handle the high density of clustered cells and their curvilinear shapes. The third stage is for building long persistent tracks using data association and track generation based on correspondence graphs \cite{Nath:ACIVS-2006}. The module involves temporal correspondence analysis, track operations such as initialization, extension, termination, recovery, and linking. Building long trajectories requires reasoning about entering and exiting cells, and track splits and merges to recover from various types of segmentation errors and accurately handle the gliding or streaming motion of cells.

\subsection{Data Analysis}

We analyzed six, four-hour-long movies using BCTracker yielding nearly 4 million bacterial instances. To generate the surface visit maps, the number of visits made by a bacterium was counted for each pixel in the field of view~\cite{Zhao2013}. A histogram of these pixel visits then comprises the visit probability $p(N)$. 

Cell reversals are marked by detecting a change in the sign of the difference of the cell velocities in two successive time points. The total number of reversals in the entire ensemble are then calculated for a given time point. Finally, the reversal frequency is calculated from the ensemble average of these reversal events over a 5 minute time window.

Number fluctuations were derived from time series data of the number of bacteria in subsystems of various sizes (from 4 $\mu m$ X 4 $\mu m$ to 110 $\mu m$ X 100 $\mu m$). From each time series, the mean number of bacteria $n$ and the standard deviation $\Delta n$ was measured. The magnitude of the number fluctuations was then quantified by the deviation in the mean number $\delta N$ and normalized by $\sqrt{n}$. For the non-reversing mutant, data from 3 movies (each 4 hours long) were used whereas data from 5 elongated streams was used for the wild type cells. 

The radial distribution function $g(r)$, was calculated from the positions $r_i$ of the bacteria using
\begin{equation*}
 g \left( r \right) = \dfrac {1}{\rho } \left\langle \sum _{i\neq 0} \delta \left( r-r_{i} \right) \right\rangle
\end{equation*}
where $\left\langle ... \right\rangle$ denotes the ensemble average.

\section{Acknowledgements} This work was supported by a Human Frontier Science Program Young Investigator Award (RGY0075/2008),  National Institutes of Health award P50GM071508, and National Science Foundation award PHY-0844466 to J.W.S; a Human Frontier Science Program Cross Disciplinary Fellowship to S.T.; and partial support for K.P. and F.B. from the Air Force Research Laboratory FA8750-14-2-0072  and NIH R33-EB00573 to K.P. The authors also wish to thank the Aspen Center for Physics where many of the ideas for this paper were developed.


\end{document}